\documentclass[lettersize,journal]{IEEEtran}
\usepackage{amsmath,amsfonts}
\usepackage{algorithmic}
\usepackage{algorithm}
\usepackage{array}
\usepackage[caption=false,font=normalsize,labelfont=sf,textfont=sf]{subfig}
\usepackage{textcomp}
\usepackage{stfloats}
\usepackage{url}
\usepackage{verbatim}
\usepackage{graphicx}
\usepackage{cite}
\usepackage{upgreek}
\usepackage{multirow}
\usepackage{bm}
\begin{document}

\title{Intrinsic Current Concentration of Buffer Layer Material for Cable Ablation Failure: Role of Random Fiber Networks}

\author{Haoran Zhang and Jianying Li$^\dag$,~\IEEEmembership{Senior Member,~IEEE}
\thanks{H. Zhang and J. Li are with the State Key Laboratory of Electrical Insulation and Power Equipment, Xi’an Jiaotong University, Xi’an 710049, China.}
\thanks{$^\dag$\,lijy@mail.xjtu.edu.cn}
}

\IEEEpubid{}

\maketitle
\begin{abstract}
In recent years, the buffer layer ablation failures of high voltage cables are frequently reported by the power systems.
Previous studies have dominantly regarded the buffer layer as the continuous homogeneous medium, whereas neglects its microstructures.
In this paper, the current distribution within the random fiber networks of buffer layer are investigated. 
Experiment results of our self-designed platform revealed an uneven current distribution in buffer layer at the moment of bearing current. 
This phenomenon is named as the intrinsic current concentration where the current density concentrates at certain sites inner the buffer layer.
And the degree of current concentration will be suppressed by compressing the sample.
Then, a 2D simulation model of the random fiber networks was constructed based on the Mikado model.
The simulation results also presented an uneven current distribution in the networks whose every fiber can be viewed as a micro-resistor.
Two types of dimensionless current concentration factors were defined to describe the degree of current concentration, finding their values decreasing with the rise of fiber density.
Meanwhile, it is equivalent of compressing the buffer layer and increasing the fiber density of model.
We believe that the intrinsic current concentration phenomenon is mainly related with the inhomogeneity of geometry structure of buffer layer.
The ablation traces and fracture fibers observed by the X-ray micro-computed tomography test supported this point.
In addition, the non-ideal surface of sample can also induce this phenomenon.
The intrinsic current concentration can aggravate the degree of originally existed macroscopic current concentration in cables, thus causing the ablation failure.
Our work may unveil a deeper understanding on the cable ablation failure and the electrical response of the similar fibrous materials.

\end{abstract}
\section{Introduction}
\begin{figure}[!t]
\centering
\includegraphics[width=3.4in]{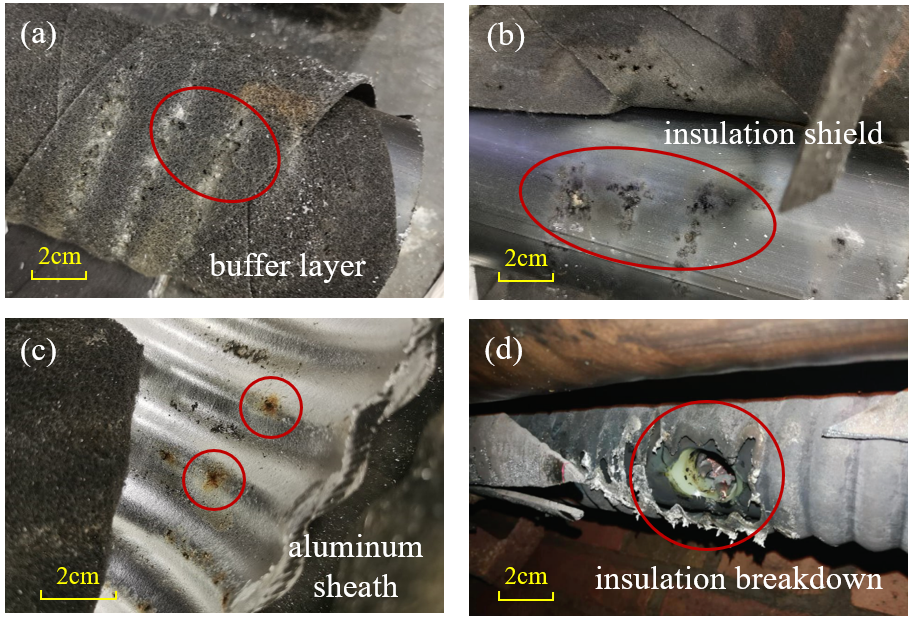}
\caption{Pictures of the typical buffer layer ablation failure. (a) ablation holes on the buffer layer, (b) burnt traces on the insulation shield, (c) burnt traces on the aluminum sheath, (d) insulation breakdown of cable.}
\label{fig1}
\end{figure}

\IEEEPARstart{H}{igh} voltage cables are the components of urban power systems, playing the essential role for electrical power transmission \cite{ref1}.
However, in recent years, there has been a notable increase in the occurrence of a specific type of buffer layer ablation failure within the high voltage cables of China, causing significant threats to the safe operation of cable systems \cite{ref2,ref3,ref4}.
At the failure sites, obvious signs of ablation and burnt marks, are observed on the buffer layer and adjacent structures \cite{ref5,ref6}.
Fig. \ref{fig1} shows the typical buffer layer ablation pictures.
These burnt traces were the evidence of once-existed high
temperature at local sites.
The ablation failure, now becoming a hot issue within the cable industry, has the potential to gradually inward deteriorate the insulation structures, ultimately culminating in the cable breakdown. 
Therefore, it is urgent to investigate the failure mechanism, and to propose the corresponding prevention methods.

In high voltage cables, the buffer layer is located between the insulating shield and the aluminum sheath. 
And one of the functions of it is to carry current along the radial direction \cite{ref7}.
Because, under the action of high voltage of the copper core, a charging current will generate and flow through the buffer layer \cite{ref8}.
This charging current is dominated by the core voltage value and the cable capacitance per unit length \cite{ref9}.
It has now been proposed by other works that the over-heat causing by the radial charging current may be an induced factor for the ablation failure.
Su \cite{ref10} reported a cable ablation failure whereas the length of extensive burnt traces was spread over 200\,m along the both side of breakdown site.
He noticed an essential fact that the number of copper strands for carrying charging current was obviously insufficient, which resulted in a poor electrical contact at the interface of buffer layer and aluminum sheath.
At this case, the current flowing through the interface will be concentrated.
In addition, Liu et al \cite{ref11} calculated the temperature rise caused by charging current concentration.
They found that the temperature value will increase with the poor electrical contact length of cable.
Under the extreme circumstances of long non-contact, the local temperature will overtake the decomposition temperature of buffer layer, thus burning the cable materials. 
Therefore, the phenomenon of current concentration may be the key to comprehend the ablation failure.

\begin{figure}[!t]
\centering
\includegraphics[width=3in]{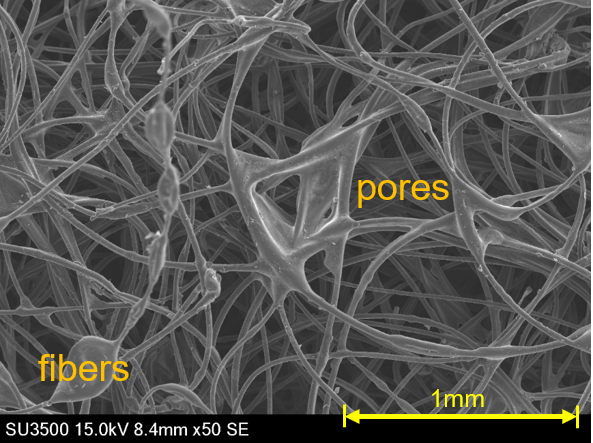}
\caption{Microscopic random fiber networks of buffer layer \cite{ref12}.}
\label{fig2}
\end{figure}

The buffer layer is composed of insulated poly(ethylene terephthalate) (PET) fibers.
In order to obtain the ability for carrying current, carbon blacks are added to PET fibers as conductive fillers during the manufacture, thus giving the buffer layer conductivity.
However, the existing researches of buffer layer ablation dominantly focuses on the macro scales, which regards the buffer layer as the continuous homogeneous medium, whereas neglects its characteristics of micro scales.
Our previous work \cite{ref12} observed the buffer layer by scanning electron microscope, revealing its fibrous microstructures comprised of PET fibers and air pores. 
Fig. \ref{fig2} presents the microscopic random fiber networks of buffer layer cited from the reference.
These type of materials with fibrous microstructures are not only existed in the topic we discussed, but also widely existed in other fields, such as flexible electronics \cite{ref13}, thermal insulation materials for aerospace \cite{ref14} and chemical catalysts \cite{ref15}.
Moreover, the random fiber networks are also universally existed in the living systems \cite{ref16}. 
For example, the intracellular scaffold known as the cytoskeleton is fibrous, which serves significant role of transporting intracellular substances and can ensure the structural integrity and mobility of cells \cite{ref17}.
Therefore, it can be significant to study the buffer layer ablation failure from the perspective of fibrous microstructures.
When the buffer layer is bearing current, every single fiber in the random networks is a micro-resistor.
And the finally current distribution is the result of the action of the overall fiber networks.
However, the current distribution inner the fiber networks of buffer layer has never been investigated, and its effects on the ablation failure are still unclear.

In this work, the current distribution in the random fiber networks of buffer layer, under the application of external current, is experimentally investigated by a self-designed platform.
Then, the numerical simulation for current distribution is conducted by finite element method (FEM).
At this part, the Mikado model is applied to construct the fiber structures of buffer layer.
Mikado model is the simplest 2D random networks, which was firstly proposed by Head et al \cite{ref18} and Wilhelm et al \cite{ref19} in 2003.
It was generated by randomly depositing lines on a 2D plane, then forming a complex network.
Subsequently, the experimental and simulation results are presented and analyzed.
Finally, this paper discusses the effects of the random fiber networks on the buffer layer ablation failure.
The findings of our work can unveil a deeper understanding on this cable failure and the relationship of the macro electrical properties and the microstructures for the similar fibrous materials.
The experimental details are described in Section \uppercase\expandafter{\romannumeral2} and the simulation model for random fibers networks is described in Section \uppercase\expandafter{\romannumeral3}.
The total results are shown in Section \uppercase\expandafter{\romannumeral4} and the related discussions are presented in Section \uppercase\expandafter{\romannumeral5}.
The conclusions are given in Section \uppercase\expandafter{\romannumeral6}.

\section{Experimental}

\subsection{Samples Preparation}
Commercial buffer layers (Shenyang Tianrong Cable Materials Co., Ltd., China) were employed as experimental samples.
These buffer layers are practically used for high voltage cables.
Before each test, the samples were dried in a cabinet for one week to thoroughly remove the internal moisture.

\subsection{Measurement of the Current Distribution in Buffer Layer}

In this part, a self-designed platform was built to measure the current distribution in buffer layer when bearing external current excitation.
It is known from the Joule's law that the heating effect of a resistor is proportional to the square of current.
Considering the current distribution can not been directly distinguished by observation, this work tries to use the surface temperature distribution of buffer layer to reflect the internal current distribution at the moment of carrying current.
As a result, the sites with higher temperature should correspond to the larger flowing current.

\begin{figure}[!t]
\centering
\includegraphics[width=3.57in]{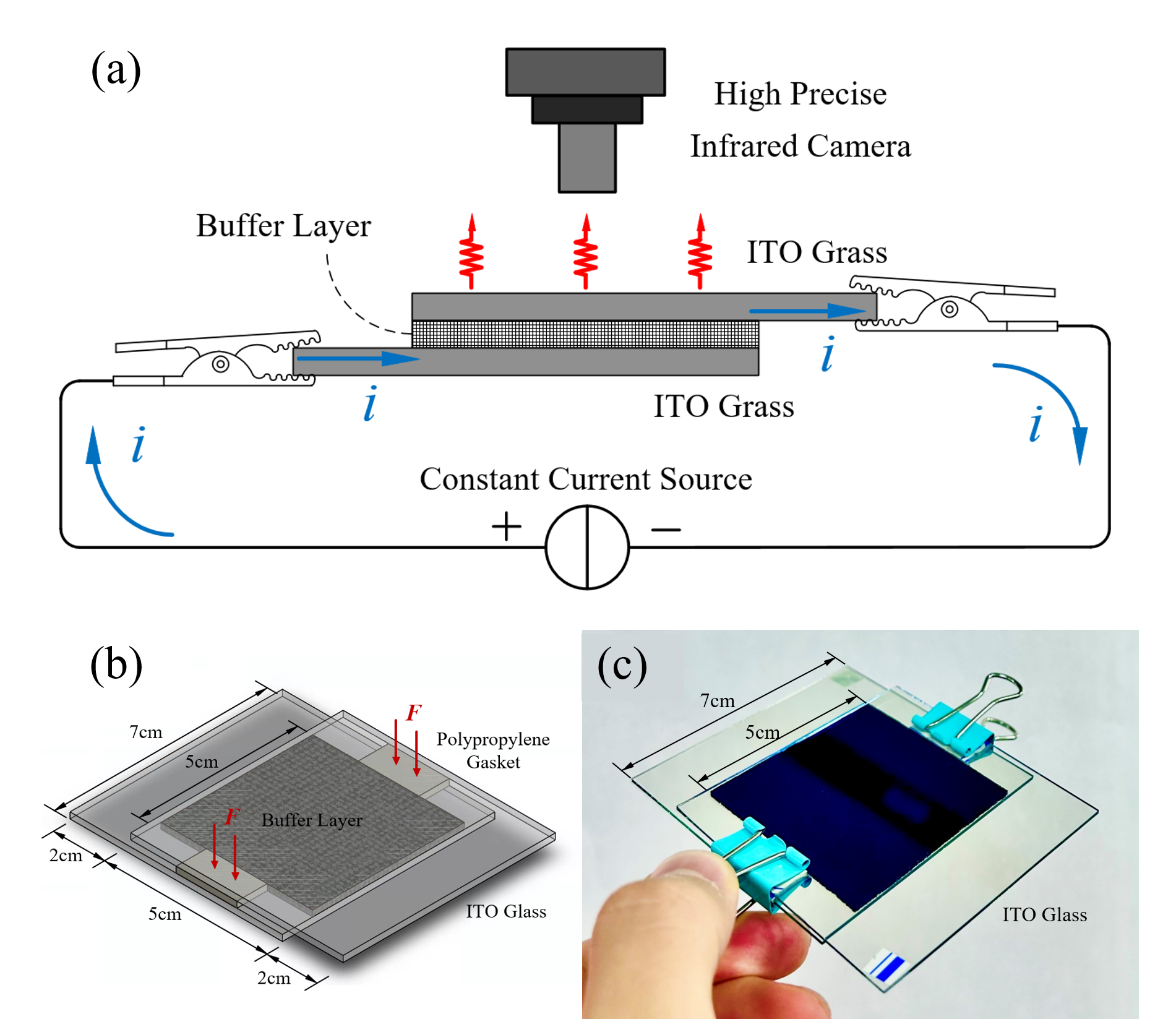}
\caption{Test platform for measuring the internal current distribution of buffer layer. (a) the design diagram, (b)$\sim$(c) the details of sample and ITO glass electrodes.}
\label{fig3}
\end{figure}

The design diagram of test platform is shown in Fig. \ref{fig3}(a), where the buffer layer sample was sandwiched between two glass electrodes, forming a three-layers structures.
These two glass electrodes were staggered, each having a terminal for power supply.
Then, a constant current generated by the power supply (VC3010A, VICTOR TECH CO., LTD, China) will vertically flow through the thickness direction of sample.
It could be seen from Fig. \ref{fig3}(a) that the current can flow the whole body of buffer layer.
On the top side of the platform, a high precise infrared camera (KD23E31, HJKIR Optoelectronic Technology CO., LTD, China) was set to record the surface temperature distribution.
This infrared camera had the ability to resolve temperature changes of micron-scale.

Fig. \ref{fig3}(b) and Fig. \ref{fig3}(c) depict the details of the buffer layer sample and electrodes.
The buffer layer was cut into a 5\,cm$\times$5\,cm square and the glass electrodes were a 7\,cm$\times$7\,cm square.
What needs additional explanation is that only one side of the glass electrode has conductivity, as a film of indium tin oxide (ITO) was deposited on the substrate.
ITO is a material combining unique transparent and conducting properties \cite{ref20}, which has been used in a wide range of applications including flexible electronics \cite{ref21}, solar cells \cite{ref22} and opto-electronic devices \cite{ref23} et al.
Compared to the traditional metal electrode, the ITO glass electrode has transparency which can provide convenience for observing sample.
And the thermal conductivity of ITO glass electrode is lower than metal, allowing the camera to capture temperature changes before the heat dissipates.
Moreover, two polypropylene gaskets were used to control the thickness of buffer layer.
The buffer layer usually has elasticity.
Under the action of external force, it will be squeezed and finally had the same thickness with gaskets.
As a result, the compression ratio $\alpha$ of buffer layer can be quantitatively defined:
\begin{equation}
   \alpha=\frac{d-d_{\mathrm{d}}}{d}\times 100\%
\end{equation}
Where $d$ is the original thickness of buffer layer, $d_{\mathrm{g}}$ is the thickness of the gaskets.

\subsection{X-Ray Micro-Computed Tomography Test}

\begin{figure}[!t]
\centering
\includegraphics[width=3.4in]{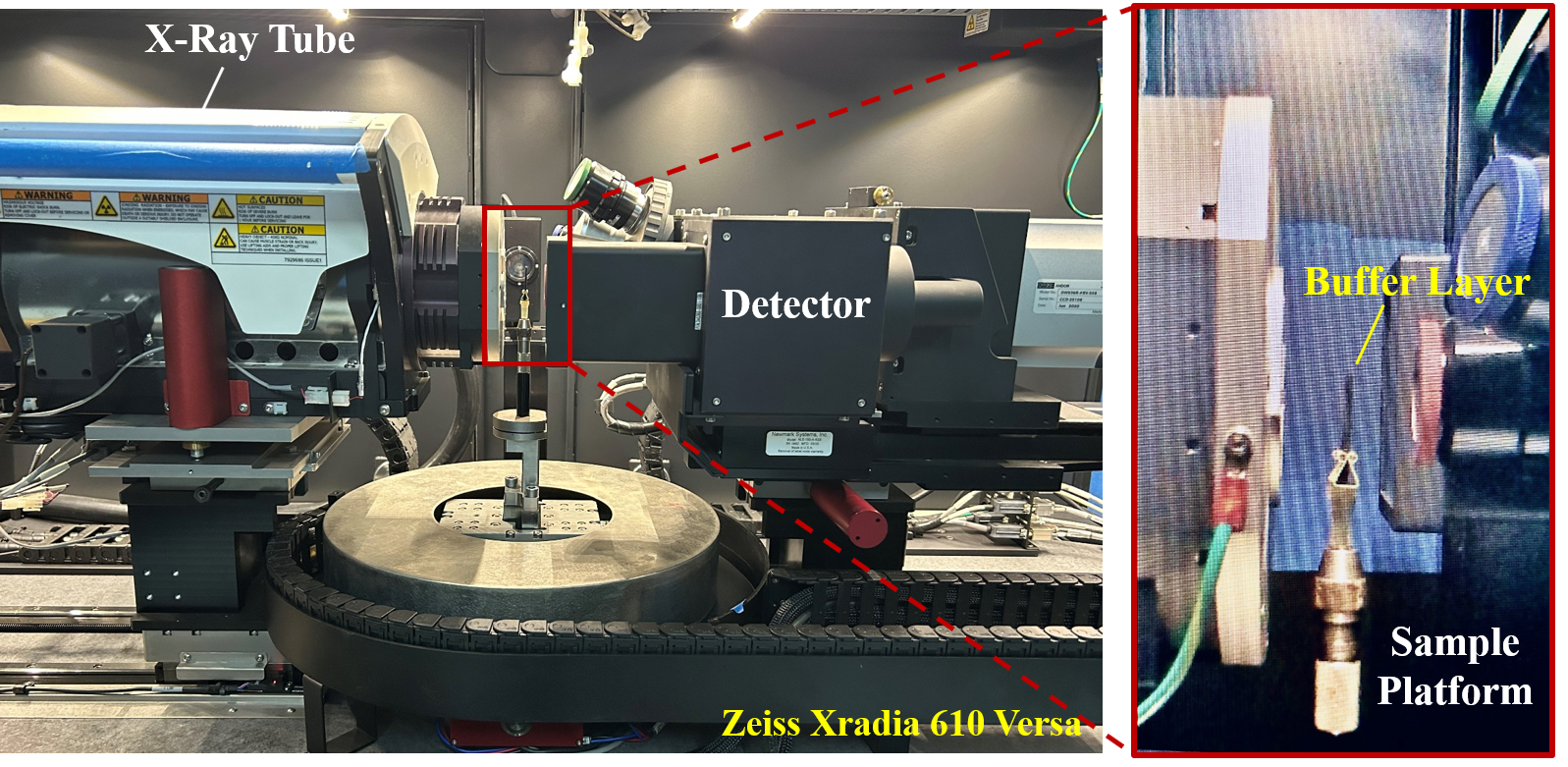}
\caption{$\upmu$CT test of the buffer layer sample.}
\label{fig4}
\end{figure}

The X-ray micro-computed tomography system can non-destructively reconstruct the 3D structures of materials \cite{ref24}. 
In this work, a high resolution system ($\upmu$CT, Zeiss Xradia 610 Versa, Germany) was employed to detect the fiber networks of buffer layer.
As presented in Fig. \ref{fig4}, the instrument consisted of a X-ray tube, a sample platform and a 2000$\times$2000 pixel detector. 
It is known that the electromagnetic X-ray wave attenuates passing through objects. 
And the attenuation coefficient is affected by the atomic number density of material. 
During the test, the transmitted X-rays were collected from different angle by rotating the sample platform, obtaining a series of 2D slices with contrasting grayscale. 
Then, the whole 3D image of sample can be reconstructed by the computed algorithm of processing software. 
The applied voltage and the power of X-ray tube is respectively 80\,kV and 35\,W. The 3D spatial resolution of test can achieve 1.8\,$\upmu$m.

\subsection{Laser Scanning Confocal Microscope Test}

The surface morphology of buffer layer was measured by a laser scanning confocal microscope system (LSCM, DCM8, Leica, Germany). 
Along the thickness direction of sample, the surface can be divided into a series of planes. 
During each scanning process, only one plane was detected. 
As the focal length of objective lens can be automatically adjusted, the series of planes were gradually scanned, finally obtaining a 3D surface morphology of sample \cite{ref25}. 
The tested light wavelength was 530\,nm, and the magnification of objective lens was 10X.


\section{Simulation Model}

Except for experimental testing, the current distribution in the random fiber networks of buffer layer are investigated by numerical simulation of FEM.
The fundamental physical laws is the current continuity equation:
\begin{equation}
    \nabla\cdot {\textbf{\textit{J}}}=0
\end{equation}
Where ${\textbf{\textit{J}}}$ is the current density. 
According to the Ohm's law ${\textbf{\textit{J}}}=\sigma{\textbf{\textit{E}}}$ and the expression of electric field ${\textbf{\textit{E}}}=-\nabla V$, the main equation solved by FEM is the Laplace's equation:
\begin{equation}
    -\nabla\cdot(\sigma\nabla V)=0
\end{equation}
Where $\sigma$ is the conductivity of material, $V$ is the electric potential. Once the potential $V$ is solved, the current distribution of model can be calculated through the negative gradient operation and multiplying conductivity.

\begin{figure}[!t]
\centering
\includegraphics[width=3.5in]{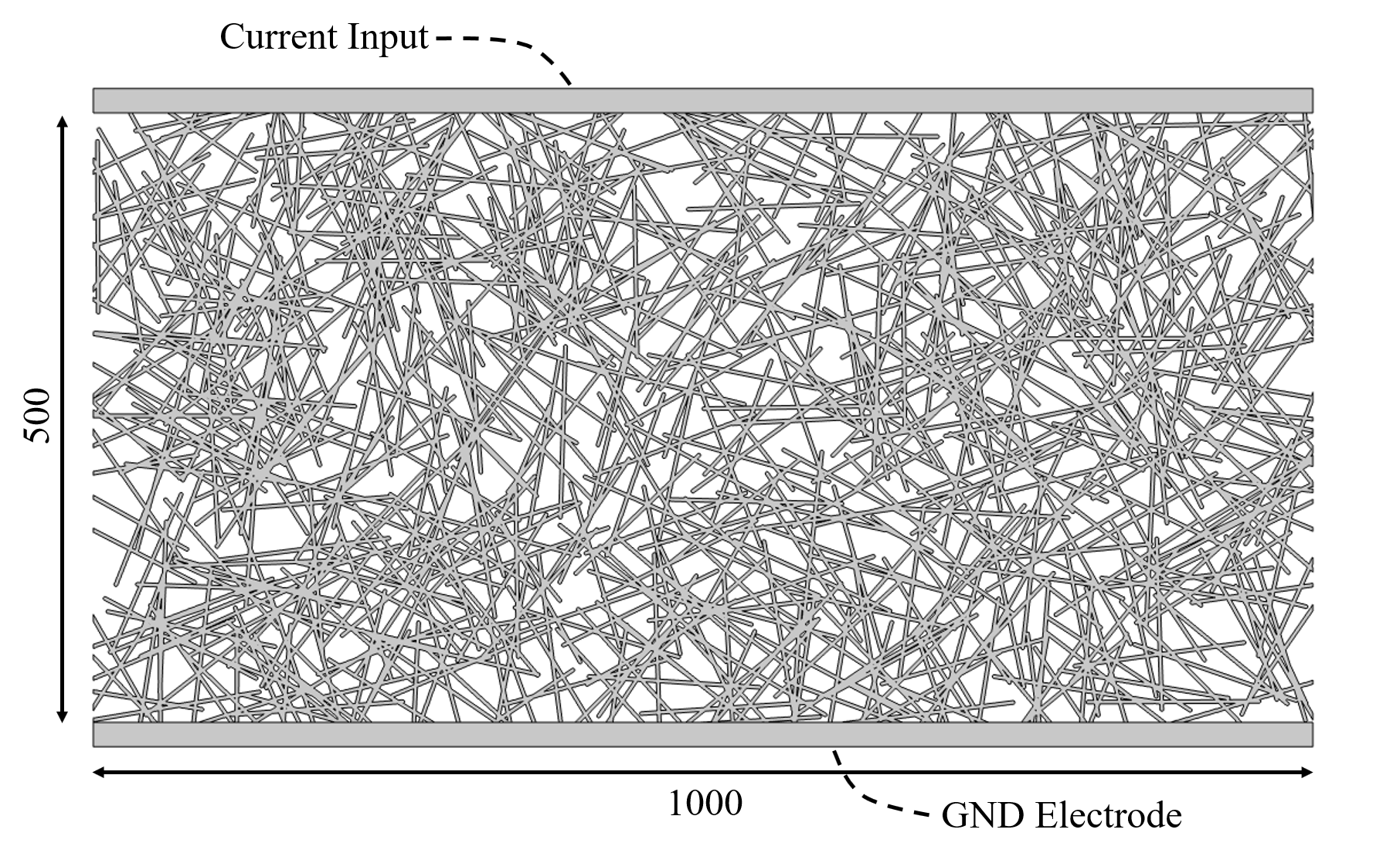}
\caption{2D simulation model based on the Mikado model for the random fiber networks of buffer layer. (\,the fiber density $\rho=0.10$\,)}
\label{fig5}
\end{figure}

Then, the key procedure of simulation is to generate the geometry of random fiber networks. 
At this part, we consider the Mikado model by randomly placing $N$ monodisperse fibers of length $l$ within a rectangular domain of area $L\times W$.
The geometry center of each fiber was randomly distributed in the rectangular area.
And the fiber orientation angle was randomly distributed in the interval $[0,\uppi]$.
Through this method, the random networks with different fiber density can be constructed.
The fiber density of networks is defined by \cite{ref19}:
\begin{equation}
    \rho=\frac{Nl}{LW}
\end{equation}
As depicted in Fig. \ref{fig5} of a simulation model of $\rho=0.10$, the geometry scale of the whole model is 1000\,$\upmu$m$\times$500\,$\upmu$m. The fiber length is set as 70\,$\upmu$m and the aspect ratio of each fiber is set as 70/3.
The conductivity of fiber is 10$^{-2}$\,S/m and the relative permittivity is set as 4741 \cite{ref12}.
Two rectangular electrodes are adjacent to the fiber networks, where the top electrode is the current input terminal with 1\,mA constant excitation and the bottom electrode is grounded.
It should be noted that our simulation models were simplified, as the fiber size and the current excitation have been reduced. 
The reason for only applying 2D model is to save the calculation time.

\section{Experimental and simulation results}

\subsection{Current Distribution in the Buffer Layer}

\begin{figure}[!t]
\centering
\includegraphics[width=2in]{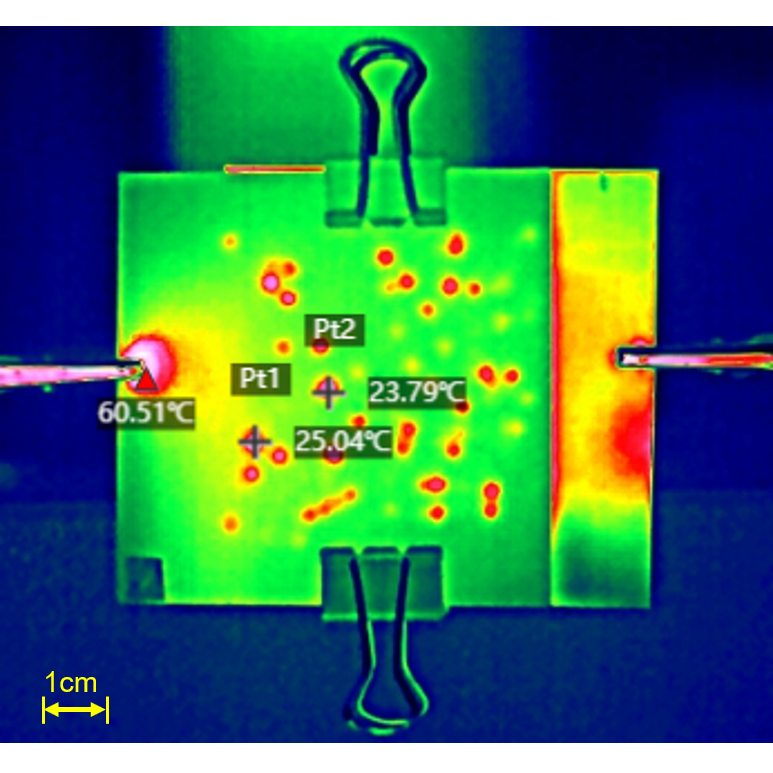}
\caption{Surface temperature distribution of the ITO glass after 5 seconds of power-on. (\,$\alpha=10\%$, $I=350$\,mA; the red spots are the hot spots representing higher temperature.\,)}
\label{fig6}
\end{figure}

In the test platform shown in Fig. \ref{fig3}, the external current was applied to buffer layer under different compression states.
The compression ratios of samples were respectively set to 10\% and 30\%.
And the surface temperatures were automatically recorded by the infrared camera in a period of time.
Since the current distribution in buffer layer cannot be directly observed, it was thus reflected by the surface temperature distribution of the ITO glass.
Fig. \ref{fig6} shows the surface temperature distribution after 5 seconds of power-on, where the external current $I$ was set to 350\,mA.
Significantly, a series of hot spots were captured by the infrared camera on surface.
The hot spots represent the regions with higher temperatures.
Fig. \ref{fig6} was shot from directly above the test platform.
From the top view, the current density in the buffer layer sample was perpendicular to the paper surface.
Therefore, the current density of the hot spots regions must be higher than those of other regions.
This reflects an important phenomenon, that is, after the buffer layer withstanding an external current, the current distribution inside the buffer layer is uneven and will be concentrated at certain sites.
In the paper, we name this phenomenon as the intrinsic current concentration phenomenon.
The so-called intrinsic refers that this phenomenon is caused by the buffer layer itself.
And our paper thinks that the phenomenon is probably related to the microstructures of the random fiber networks of buffer layer.

\begin{figure}[!t]
\centering
\includegraphics[width=3.4in]{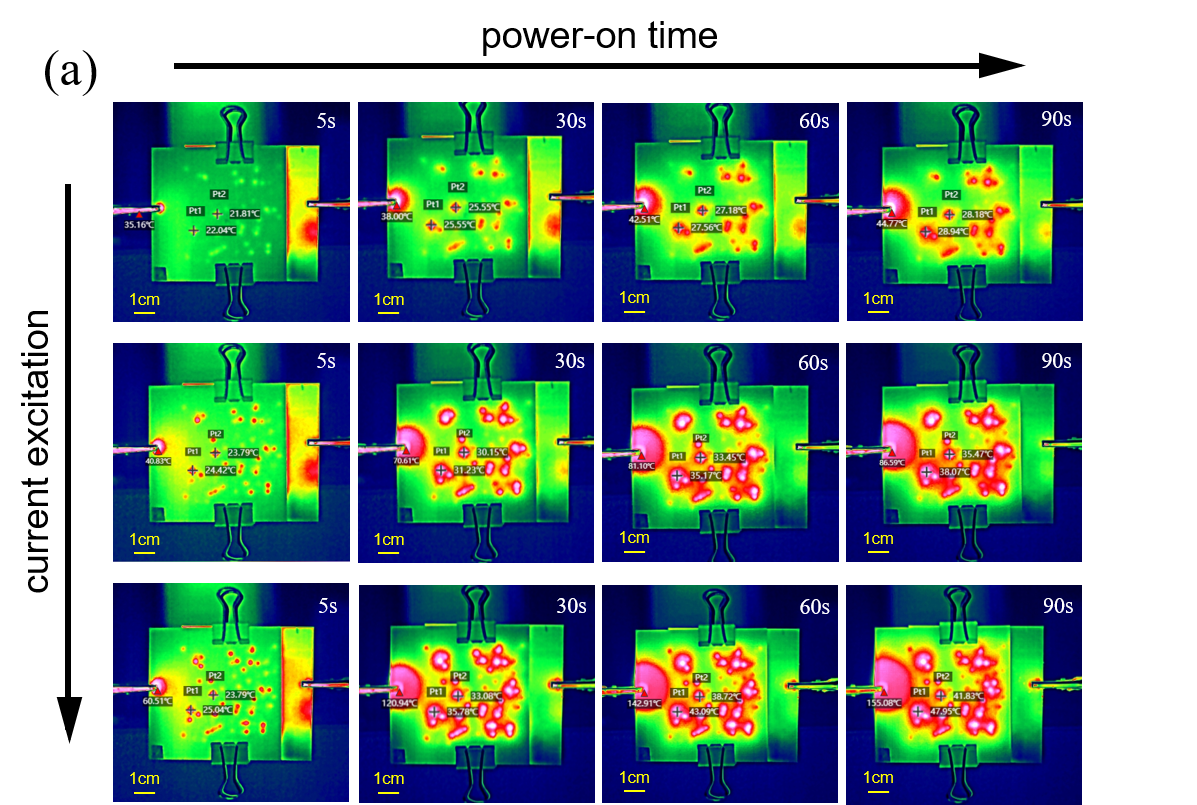}
\includegraphics[width=3.4in]{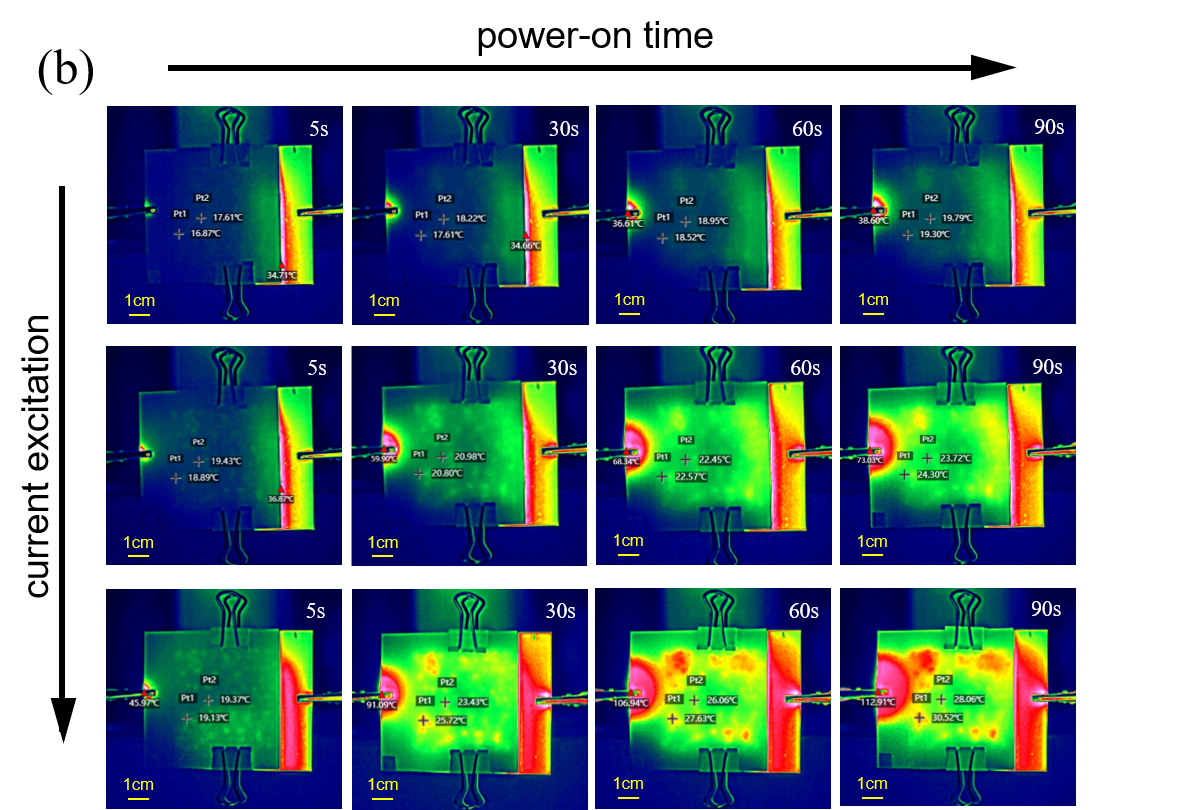}
\caption{Intrinsic current concentration phenomenon of buffer layer under different experimental conditions. (a) $\alpha=10\%$, (b) $\alpha=30\%$.}
\label{fig7}
\end{figure}

The intrinsic current concentration phenomenon obtained under different experimental conditions are depicted in Fig. \ref{fig7}.
The power-on time of the pictures gradually prolonged from left to right, and the amplitude of external current gradually increased from top to bottom.
Through observing Fig. \ref{fig7}(a) and Fig. \ref{fig7}(b), the area of hot spots obviously expands along the increase of external current.
It means that the degree of current concentration aggravates.
Moreover, it is found that the intrinsic current concentration can be affected by the compression ratio of buffer layer.
When the compression ratio $\alpha$ increases from 10\% to 30\%, the area and number of the previously existing hot spots have been significantly reduced.
It indicates that compressing the buffer layer along the thickness direction can improve the evenness of current distribution inside the material.
Table \ref{tab1} shows the minimum temperature $T_{\mathrm{min}}$, the average temperature $T_{\mathrm{avg}}$ and the maximum temperature $T_{\mathrm{max}}$ of the ITO glass surface during the experiments.
The $T_{\mathrm{avg}}$ lies exactly between the $T_{\mathrm{min}}$ and the $T_{\mathrm{max}}$.
As the power-on time and external current increase, each temperature values display an upward tendency.
It is also seen that increasing the compression ratio can decrease the surface temperature values, which indicates the current density of the hot spots reduced.
These are consistent with the observation results from Fig. \ref{fig7}.

\begin{table*}[!t]
\caption{Recorded temperature values during the experiments\label{tab1}}
\renewcommand\arraystretch{1.5}
\centering
\begin{tabular}{|c|c|c|c|c|c||c|c|c|c|}
\hline
        ~ & ~ & \multicolumn{4}{c||}{$\bm{\alpha=10\%}$} & \multicolumn{4}{c|}{$\bm{\alpha=30\%}$} \\ \hline
        ~ & ~ & $t=5\,\mathrm{s}$ & $t=30\,\mathrm{s}$ & $t=60\,\mathrm{s}$ & $t=90\,\mathrm{s}$ & $t=5\,\mathrm{s}$ & $t=30\,\mathrm{s}$ & $t=60\,\mathrm{s}$ & $t=90\,\mathrm{s}$   \\ \hline
        \multirow{2}*{$\bm{T_{\mathrm{min}}}(^\circ \mathrm{C})$} & $I=150\,\mathrm{mA}$ & 17.86 & 18.28 & 19.01 & 19.55 & 16.56 & 17.05 & 17.61 & 18.09  \\ \cline{2-10}
        ~                                       & {$I=350\,\mathrm{mA}$} & 18.58 & 20.21 & 22.80 & 25.27 & 17.48 & 19.79 & 21.46 & 22.97  \\ \hline
        \multirow{2}*{$\bm{T_{\mathrm{avg}}}(^\circ \mathrm{C})$} & $I=150\,\mathrm{mA}$ & 19.54 & 21.58 & 23.44 & 24.51 & 18.62 & 19.38 & 20.11 & 20.86  \\ \cline{2-10}
        ~                                       & {$I=350\,\mathrm{mA}$} & 21.14 & 27.96& 33.98 & 38.01 & 19.80 & 23.05 & 26.03 & 28.33  \\ \hline
        \multirow{2}*{$\bm{T_{\mathrm{max}}}(^\circ \mathrm{C})$}   & $I=150\,\mathrm{mA}$ & 22.33 & 25.67 & 27.85 & 29.22 & 21.57 & 22.04 & 22.68 & 23.56  \\ \cline{2-10}
        ~                                       & {$I=350\,\mathrm{mA}$} & 25.27 & 36.09 & 43.38 & 47.99 & 21.69 & 25.78 & 30.37 & 33.45  \\ \hline
\end{tabular}
\end{table*}

\subsection{Characterization Results of the Random Fiber Networks by $\mu$CT}

\begin{figure}[!t]
\centering
\includegraphics[width=3.4in]{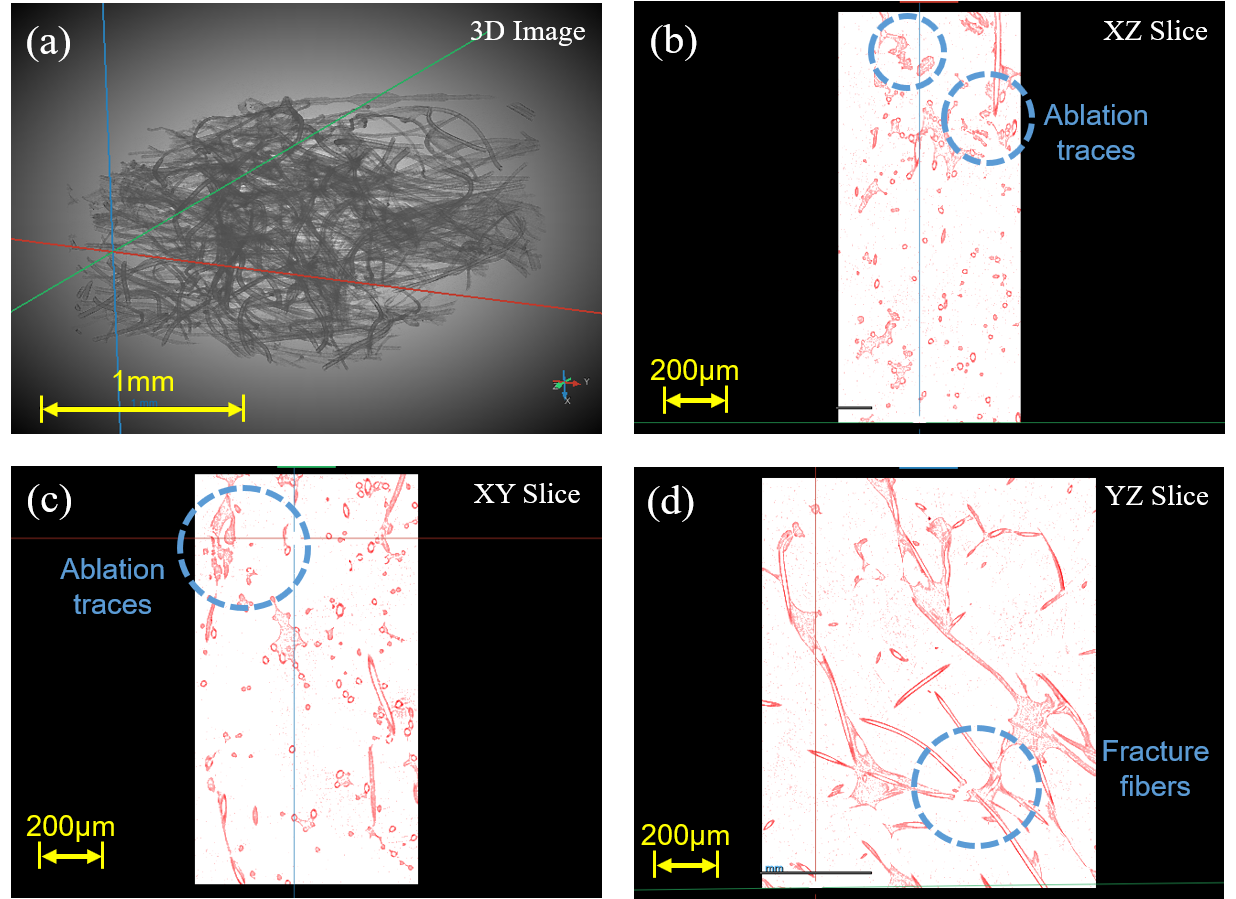}
\caption{$\upmu$CT results of an ablation buffer layer sample. (a) 3D image of the fiber networks, (b) XZ slice, (c) XY slice, (d) YZ slice.}
\label{fig8}
\end{figure}

The random fiber networks of buffer layer were characterized by µCT.
In the experiment, an ablation buffer layer sample which had ever suffered from large flowing current was tested, and Fig. \ref{fig8} presents its $\upmu\mathrm{CT}$ results.
The original µCT data was the images of grayscale.
Then, the regions of interests were divided by manually setting grayscale range.
In Fig. \ref{fig8}(a), the skeleton of fiber networks were reconstructed and the orientation of each fiber can be clearly recognized.
In the 2D slices of Fig. \ref{fig8}(b) to Fig. \ref{fig8}(d), the fibers were rendered as red areas.
It could be seen that there are several ablation traces on these 2D slices, where the integrity of fiber networks were destroyed.
Especially, the evidence of fracture fibers was clearly shown in Fig. \ref{fig8}(d).
It indicates that the destroy of larger current for buffer layer, is actually reflected as the microscopic destroy for the fiber networks.
And the fiber networks intrinsically have some weaknesses.
Under the excitation of external current, the current density at the weakness sites will exceed the tolerance threshold of fibers, which then damages buffer layer.

\subsection{Current Distribution in the 2D Simulation Model}

\begin{figure}[!t]
\centering
\includegraphics[width=3.4in]{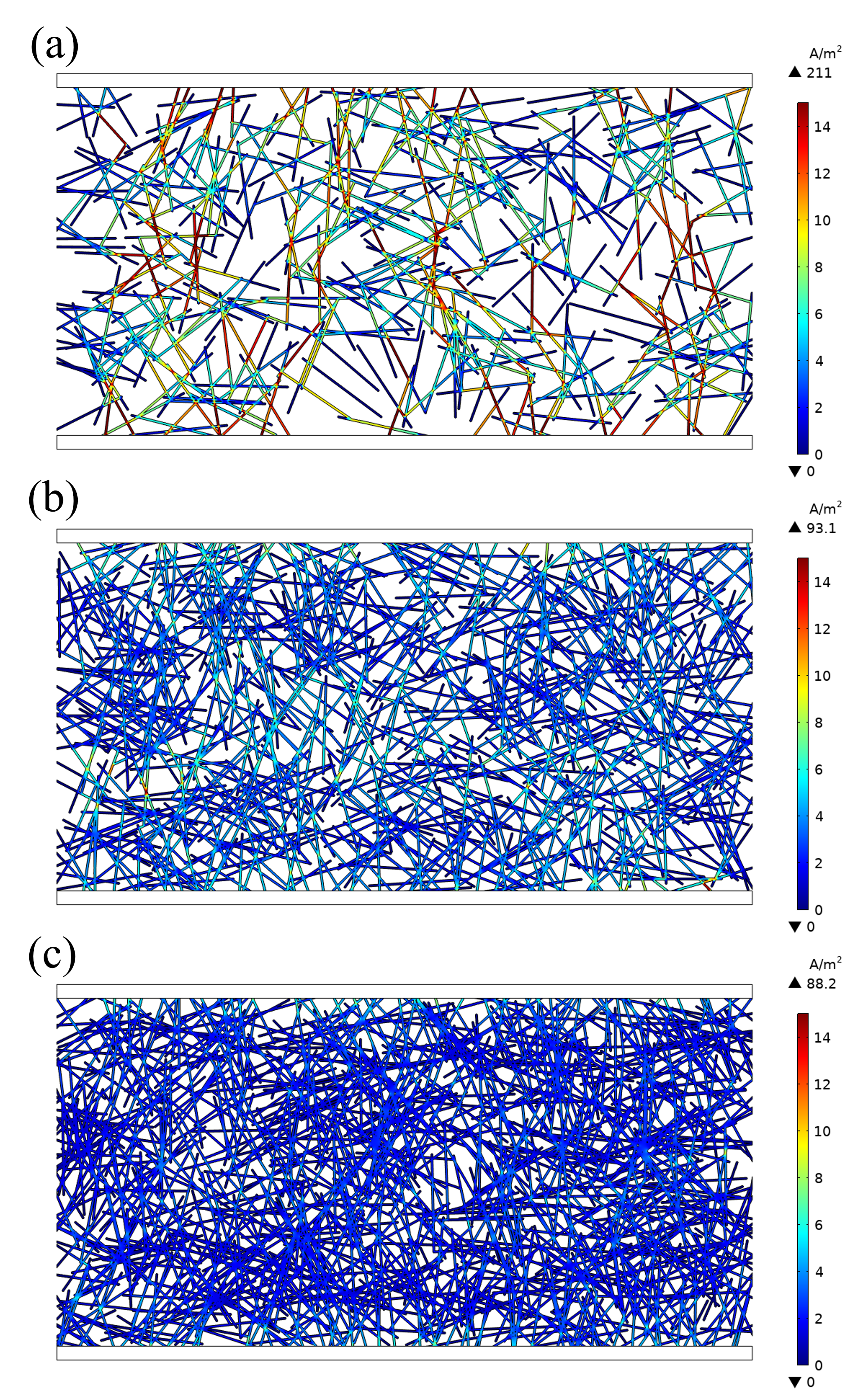}
\caption{Simulation results of the current distribution in the 2D simulation model. (a) $\rho=0.05$, (b) $\rho=0.10$, (c) $\rho=0.15$. (\,All the calculations has the same applied current excitation.\,)}
\label{fig9}
\end{figure}

In order to investigate the relationships between the intrinsic current concentration and the random fiber networks, the current distribution inner the 2D simulation model, which was constructed based on the Mikado model, was calculated in this part.
Fig. \ref{fig9} shows the results of different fiber density.
All the calculations has the same applied current excitation.
It can be seen that the current distribution in the random fiber networks is uneven, where some fibers will withstand more current density than others.
As presented in Fig. \ref{fig9}(a), the degree of current unevenness is more severe for lower density networks. 
In this case, the current will be concentrated on several fibers or the cross-linked regions of the fibers.
It is conceivable that these sites will be priorly destroyed under the external current and they are the weaknesses of the fiber networks.
Therefore, the simulation results can support the $\upmu$CT results of weaknesses shown in Fig. \ref{fig8}.

\begin{figure}[!t]
\centering
\includegraphics[width=3.4in]{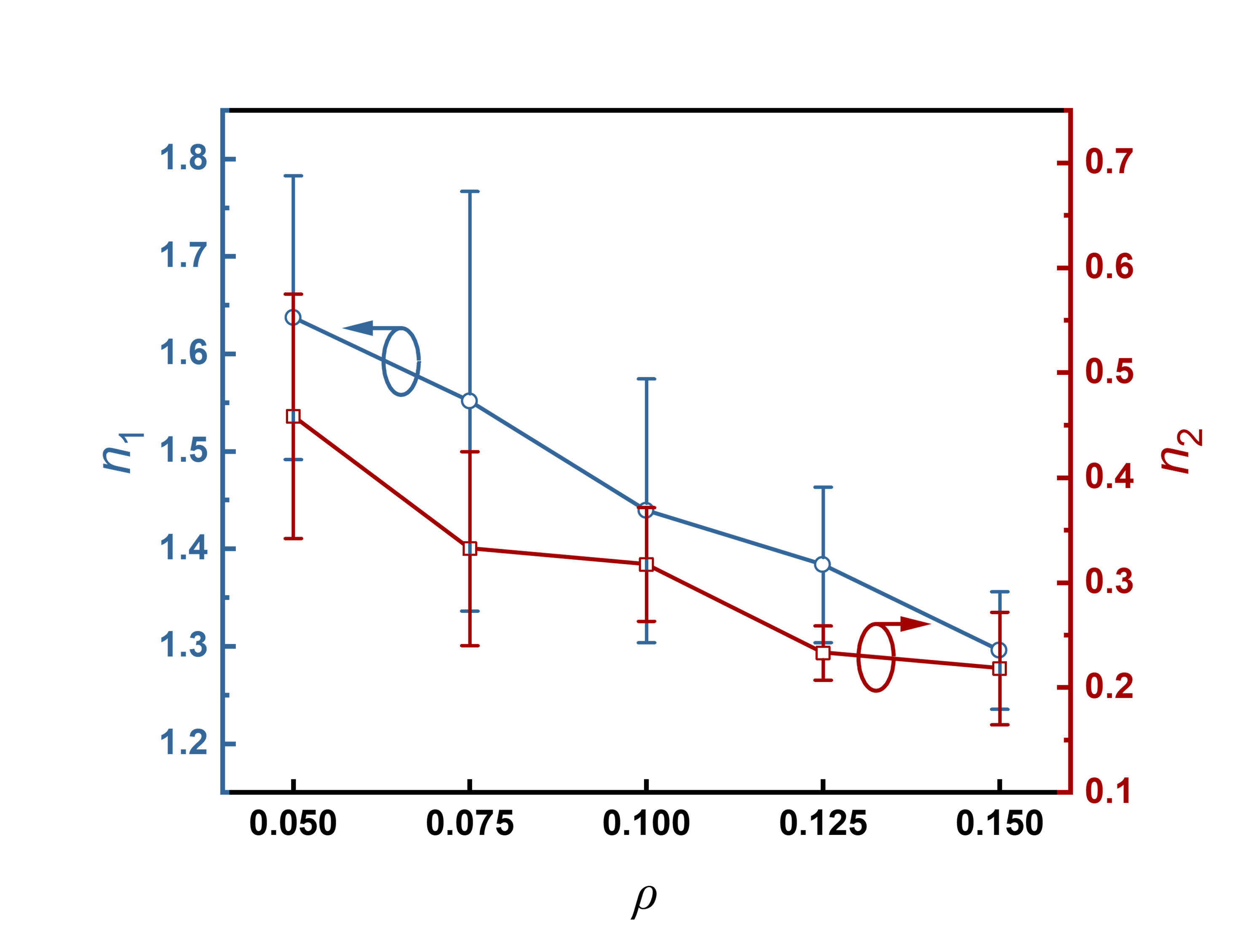}
\caption{Dimensionless current concentration factors $n_1$ and $n_2$ with the fiber density $\rho$.}
\label{fig10}
\end{figure}

Furthermore, the degree of current concentration in the networks can be quantitatively described by two type of dimensionless current concentration factors ($n_1$ and $n_2$), which are defined by referring to the related work of Topcagic \cite{ref26}. 
\begin{equation}
    I_{i}=\int_{l_{i}}\,{\textbf{\textit{J}}}\cdot{\textbf{{n}}}\,\mathrm{d}l\quad(i=1,2,\,\cdots,10)
\end{equation}
\begin{equation}
    n_{1}=\frac{{(I_{i})}_{\mathrm{max}}}{I_{\mathrm{avg}}}
\end{equation}
\begin{equation}
    n_{2}=\frac{\sqrt{\frac{\sum_{i=1}^{10}{(I_{i}-I_{\mathrm{avg}})}^2}{10}}}{I_{\mathrm{avg}}}
\end{equation}
where the upper electrode of the model are evenly divided into ten parts along the length. The current flowing through each part of electrode $I_{i}$ is the line flux of the current density $\textbf{\textit{J}}$ along the normal direction $\textbf{{n}}$. The current concentration factor $n_1$ is the ratio of the maximum electrode current to the average electrode current.
And the current concentration factor $n_2$ is the ratio of the standard deviation of the electrode current to the average electrode current.
Obviously, as the degree of current concentration aggravates, the two types of current concentration factors will increase.
Fig. \ref{fig10} presents the curves of current concentration factors with the fiber density.
For every fiber density value, five random fiber networks were constructed, and their current concentration factors were calculated according to Eq. (6) and Eq. (7).
Therefore, the data points of Fig. \ref{fig10} are added with the error bars.
It is revealed that the current concentration factors will decrease with the increasing of fiber density within a certain range.
This simulation result can perfectly explain the aforementioned experiment phenomena, where the intrinsic current concentration can be suppressed by compressing buffer layer.
Because according to the definition of fiber density of Eq. (4), compressing buffer layer will not change the number of fibers, but it will decrease the width of the model, thus reducing the denominator of Eq. (4) and leading to the rise of fiber density.
Therefore, it is equivalent of compressing buffer layer and
increasing the fiber density value.

\subsection{Surface Morphology of the Buffer Layer}

\begin{figure}[!t]
\centering
\includegraphics[width=3.4in]{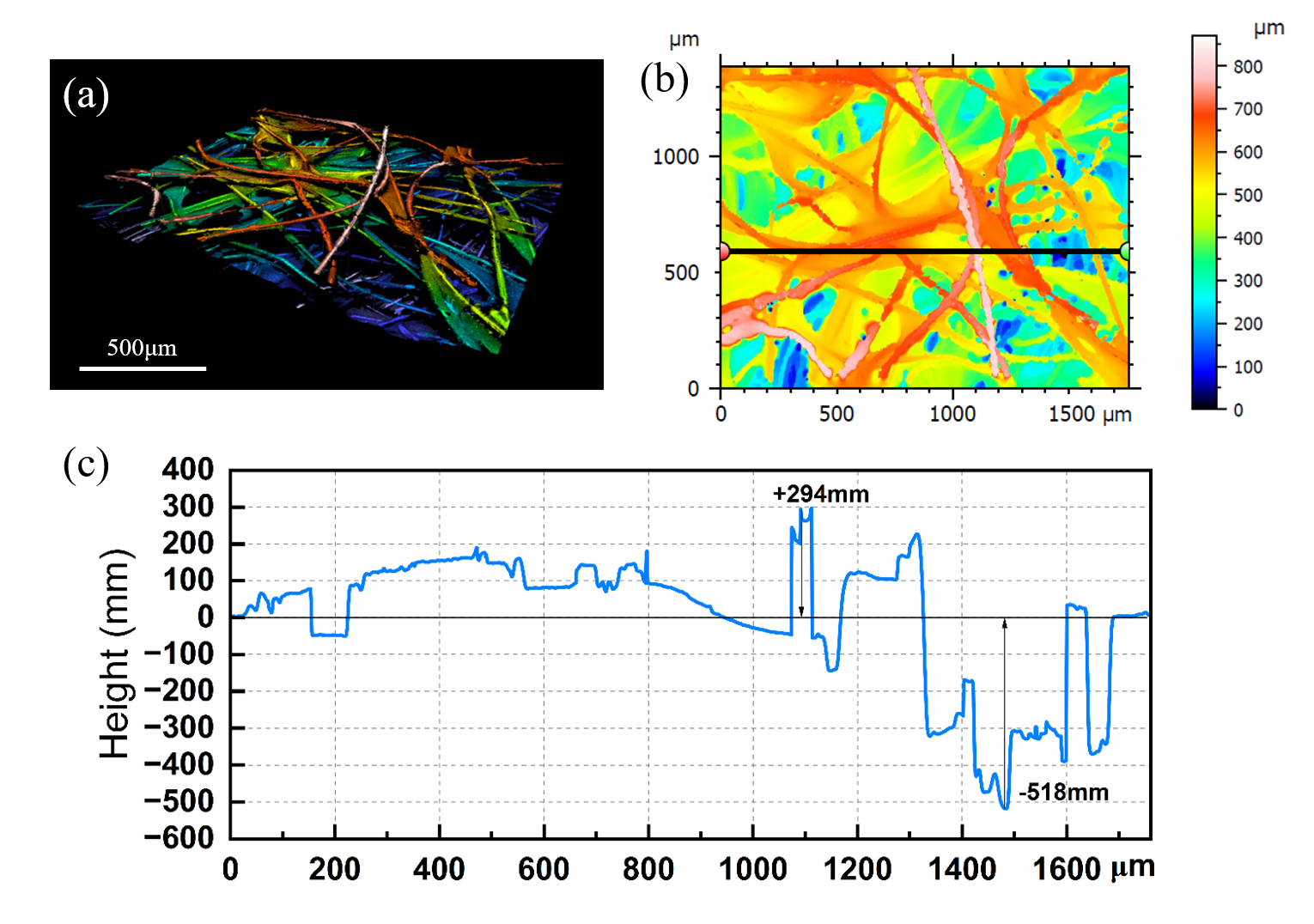}
\caption{Surface morphology of the buffer layer. (a) 3D image, (b) 2D image, (c) line roughness result.}
\label{fig11}
\end{figure}

Fig. \ref{fig11} shows the surface morphology of buffer layer tested
by LSCM. 
It is clear that the surface of buffer layer was not ideal, which has a height difference.
As presented in Fig. \ref{fig11}(a) and Fig. \ref{fig11}(b), the red area was the higher site, whereas the blue area was the lower site. 
In order to measure the fluctuations of surface, a line roughness was drawn at the selected black line in Fig. \ref{fig11}(b), and the result is presented in Fig. \ref{fig11}(c).
It can be seen that the highest height of the line was +294\,mm and the lowest value is -518\,mm.
Therefore, there is height difference over 800\,mm along the selected line.
Based on the aforementioned results, the rough surface morphology observed in the buffer layer can be attributed to its fibrous microstructures, as the buffer layer is not a continuous homogeneous medium.

\section{Discussions}

\subsection{Reasons for the Intrinsic Current Concentration}

Every fiber of the buffer layer is a micro-resistor, serving as the path for current flow.
The intrinsic current concentration phenomenon reveals that current distribution in the fiber networks is uneven, which will concentrate at the certain sites.
It is believed that the current concentration dominantly results from the inhomogeneity of the geometry structure of the fiber networks.
Because, the simulation results of the randomly generated networks of Fig. \ref{fig9} demonstrate this phenomenon.
And the degree of current concentration will be affected by the compression state of buffer layer.
As the fiber density increases, the homogeneity of the networks geometry is improved and the current concentration will be suppressed.

Moreover, another factor for intrinsic current concentration may be related with the surface morphology of buffer layer.
According to the results of LSCM, the surface of buffer layer were not ideal, having the height difference of hundreds of microns. 
Therefore, when the buffer layer is contacted with the electrode, only the fibers of higher sites can actually contacted with electrode, and the fibers of lower sites will be blocked by the air gaps.
As a result, the external current will merely flow through these contacted sites.

\subsection{Effects of the Intrinsic Current Concentration on the Buffer Layer Ablation Failure}

\begin{figure}[!t]
\centering
\includegraphics[width=3.4in]{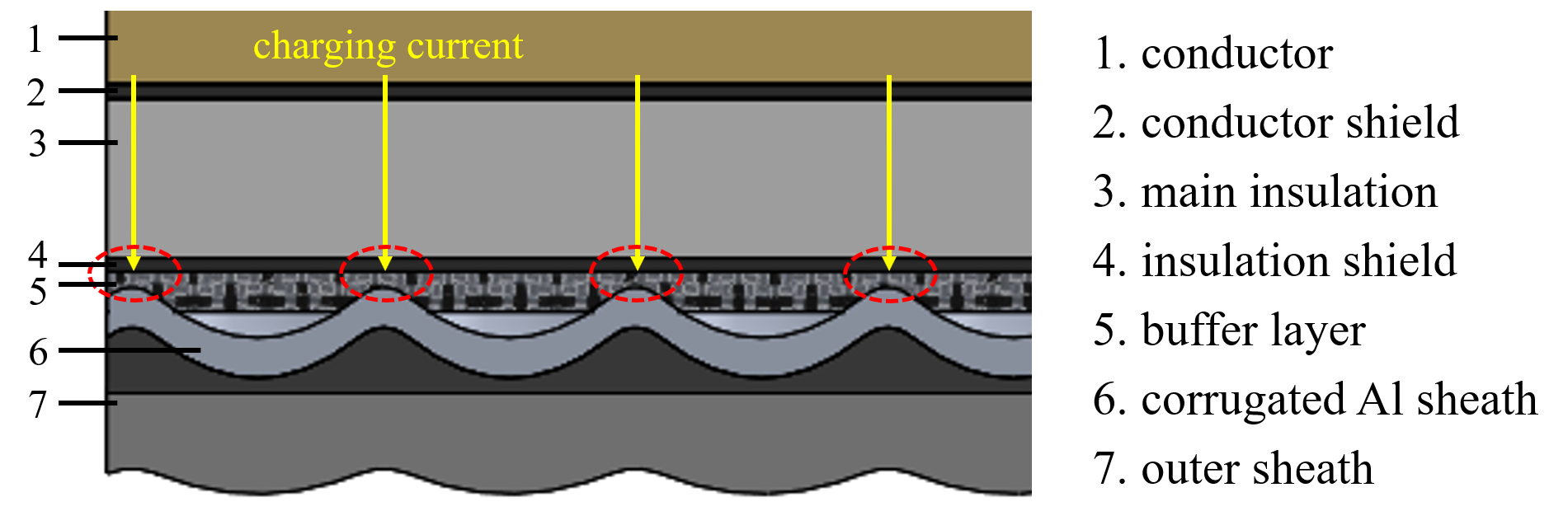}
\caption{Schematic diagram of the cable structure. (\,The
charging current was described as the yellow arrows.\,)}
\label{fig12}
\end{figure}

As presented in Fig. \ref{fig12}, the buffer layer plays the role of carrying the charging current within the cables.
The charging current is generated from the high voltage conductor, and will then flow through the conductor shield, the main insulation, the insulation shield, and finally pass through the buffer layer into the corrugated aluminum sheath.
Due to the application of corrugated aluminum structure, the buffer layer can form the electrical contact with the troughs of aluminum sheath.
Therefore, there is an originally existed macroscopic current concentration phenomenon in high voltage cables, where the charging current can merely flow through those contacted sites as shown by the red circle marks in Fig. \ref{fig12}.
Our paper finds the intrinsic current concentration phenomenon of buffer layer caused by the inhomogeneity of microscopic fiber networks.
Obviously, this type of microscopic current concentration will aggravate the degree of the existed macroscopic current concentration, thus increasing the risk of ablation failure.
Therefore, it is of considerable significance to focus on the microscopic fiber networks of buffer layer during the manufacture and application of cables.
And further methods should be proposed to improve the evenness of current distribution in buffer layer and weaken this intrinsic current concentration phenomenon.

\section{Conclusions}
In this article, the current distribution in the fiber networks of buffer layer was investigated through experiments and simulations.
The reasons for the current distribution characteristics and their effects on the buffer layer ablation failure were discussed.
The main conclusions are summarized as follows:

\begin{enumerate}
\item{A series of hot spots appeared on the surface of buffer layer at the moment of bearing current, which means the current distribution within the buffer layer was uneven. Our paper names this phenomenon as intrinsic current concentration. The word of intrinsic indicates the phenomenon was caused by buffer layer itself. And the degree of intrinsic current concentration will be affected by the buffer layer compression ratio and external current amplitude. As the buffer layer compression ratio increases and the external current decreases, the intrinsic current concentration phenomenon will be suppressed.}
\item{The 2D simulation model for the random fiber networks of buffer layer was constructed based on the Mikado model. Every fiber in the networks is a micro-resistor and the current distribution within the whole networks is also uneven. The intrinsic current concentration phenomenon dominantly results from the inhomogeneity of the geometry structure of buffer layer. The ablation traces and fracture fibers also observed by the $\upmu$CT test of the networks supported this point. Two types of dimensionless current concentration factors were proposed to quantitatively described current concentration degree, finding the equivalence of compressing buffer layer and increasing the fiber density.
In addition, the intrinsic current concentration phenomenon is also related with the non-ideal surface of buffer layer, which was confirmed by the LSCM test.}
\item{There is an originally existed macroscopic current concentration  in high voltage cables due to the corrugated structures of aluminum sheath. The intrinsic current concentration will aggravate the degree of macroscopic current concentration in cables, thus increasing the risk of ablation occurrence. Therefore, the homogeneity of buffer layer should be improved to reduce the risk of current concentration.
}
\end{enumerate}

Our work can deepen the understanding of mechanism of buffer layer ablation and the electrical response of similar fibrous materials.
The further work could focus on optimizing these random fiber networks to reduce the danger for cables operation.

\section*{Acknowledgments}
The authors thank Ms. Guo and Mr. Ren at Instrument Analysis Center of Xi’an Jiaotong University for their assistance with LSCM and $\upmu$CT analysis. The authors also thank the colleagues H. Sui and B. Li for their assistance with experiments.

\begin{IEEEbiography}[{\includegraphics[width=1in,height=1.25in,clip,keepaspectratio]{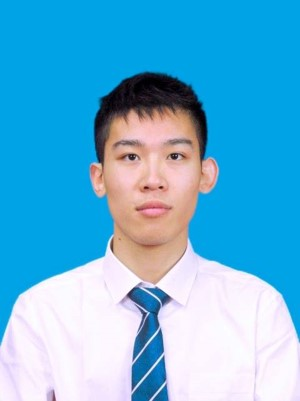}}]{Haoran Zhang}
was born in Anhui, China, in 1998. He received the B.S. and M.S. degree in electrical engineering from Xi'an Jiaotong University, Xi'an China, in 2020 and 2024 respectively.
His research interest is engineering or applied physics.

\end{IEEEbiography}

\vspace{11pt}

\begin{IEEEbiography}[{\includegraphics[width=1in,height=1.25in,clip,keepaspectratio]{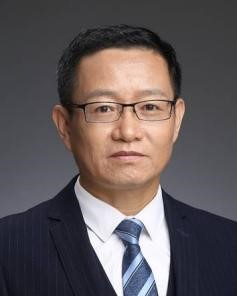}}]{Jianying Li}
(Senior Member, IEEE) was born in Shaanxi, China, in 1972. He received the B.S., M.S., and Ph.D. degrees in electrical engineering from Xi'an Jiaotong University, Xi'an China, in 1993, 1996, and 1999, respectively.

He is currently a Professor in the State Key Laboratory of Electrical Insulation and Power Equipment, Xi'an Jiaotong University.
His major research interests include advanced electrical materials and dielectric physics.
\end{IEEEbiography}

\vfill

\end{document}